\documentclass[preprint,12pt]{elsarticle}

\usepackage{latexsym,amsmath,amssymb}

\newcounter{bla}

\journal{Computer Physics Communications}

\begin{document}

\begin{frontmatter}



\title{MESAFace, a graphical interface to analyze the MESA output}


\author{M. Giannotti\corref{M. Giannotti}}
\author{M. Wise}
\author{A. Mohammed}

\cortext[M. Giannotti] {Corresponding author.\\\textit{E-mail address:} mgiannotti@mail.barry.edu}
\address{Barry University,
11300 NE 2nd Ave., Miami Shores, FL 33161, US}

\begin{abstract}
MESA (Modules for Experiments in Stellar Astrophysics) has become very popular among astrophysicists as a powerful and reliable code to simulate stellar evolution. 
Analyzing the output data thoroughly may, however, present some challenges and be rather time-consuming. 

Here we describe MESAFace, a graphical and dynamical interface which provides an intuitive, efficient and quick way to analyze the MESA output.
\end{abstract}


\end{frontmatter}



{\bf PROGRAM SUMMARY/NEW VERSION PROGRAM SUMMARY}

\begin{small}
\noindent
{\em Manuscript Title:} MESAFace, a graphical interface to analyze the MESA output                                       \\
{\em Authors:} M. Giannotti, M. Wise, A. Mohammed                                                \\
{\em Program Title:} MESAFace\\
{\em Journal Reference:}                                      \\
{\em Catalogue identifier:}                                   \\
{\em Licensing provisions:} none                                   \\
{\em Programming language:} Mathematica                                \\
{\em Computer:} Any computer capable of running Mathematica.                                               \\
{\em Operating system:} Any capable of running Mathematica.
Tested on Linux, Mac, Windows XP, Windows 7.                                       \\
{\em RAM:} Recommended 2 Gigabytes or more.                                              
{\em Classification:}   1.7 Stars and Stellar Systems, 14 Graphics                                  \\
{\em Nature of problem:} \\
Find a way to quickly and thoroughly analyze the output of a MESA run, including all the profiles, and have an efficient method to produce graphical representations of the data.
   \\
{\em Solution method:} \\
We created two scripts (to be run consecutively).
The first one downloads all the data from a MESA run and organizes the profiles in order of age.
All the files are saved as tables or arrays of tables which can then be accessed very quickly by Mathematica.
The second script uses the \texttt{Manipulate} function to create a graphical interface which allows the user to choose what to plot from a set of menus and buttons. 
The information shown is updated in real time. 
The user can access very quickly all the data from the run under examination and visualize it with plots and tables.   \\
{\em Unusual features:}\\
Moving the slides in certain regions may cause an error message. This happens when Mathematica is asked to read nonexistent  data.
The error message, however, disappears when the slides are moved back. This issue does not preclude the good functioning of the interface.     
   \\
{\em Additional comments:}\\ 
The program uses the dynamical capabilities of Mathematica.
When the program is opened, Mathematica prompts the user to "Enable Dynamics". 
It is necessary to accept before proceeding.
\\
{\em Running time:} \\
Depends on the size of the data downloaded, on where the data are stored (hard-drive or web), and on the speed of the computer or network connection.
In general, downloading the data may take from a minute to several minutes. 
Loading directly from the web is slower.
For example, downloading a 200MB data folder (a total of 102 files) with a  dual-core Intel laptop, P8700, 2 GB of RAM,  at 2.53 GHz took about a minute from the hard-drive and about 23 minutes from the web (with a basic home wireless connection).   \\

\end{small}
%

\section{Introduction}
\label{sec:Introduction}

The introduction of the new code for stellar evolution, MESA~\cite{Paxton:2010ji} (Modules for Experiments
in Stellar Astrophysics) represented somewhat of a revolution in the field of computational stellar evolution.

MESA is a one-dimensional stellar evolution code, organized in independent modules and continually updated.
The code can evolve stars with a very wide range of initial masses and metallicities; it allows the specification of many parameters, for example, convection mechanism, mass loss, etc., and it can overcome difficulties typical of the previous stellar evolutionary codes. 
Notably, it can run a low mass star through the He-flash region.

All these aspects and the fact that the code is publicly available~\cite{mesa} has made MESA extremely popular among astrophysicists and other physicists interested in stellar evolution and in how stars respond to new models of particle interactions.

Being publicly available, MESA is widely accessible and is continually tested by hundreds of scientists and students all over the world.
In addition, several researchers have created new codes and routines to improve it and to simplify its use~\cite{tools}.

A MESA run produces a very large amount of data on the life and structure of the star.
This information is saved in several files:
the file "star.log" includes global information about the star at different times whereas a large number of \textit{profiles}
(named "log$n$.data", where $n$ is an integer), contain detailed information about the profile of the star, each file referring to a specific age.

Though it is relatively easy, and for some purposes sufficient, to analyze the file "star.log", 
a complete analysis of the MESA output, which includes in some cases hundreds of  profiles, is much more challenging. 
In particular, to analyze the detailed structure of a star at a certain age, 
one should select the corresponding profile by reading the age of each of them (the information is provided in the $3^{\rm nd}$ row of "log$n$.data").
Alternatively, one should read the \textit{model number} corresponding to a certain age in the first column of the file "star.log", and then look for the log number associated with that age in the file "profile.index".
Both ways are long and, in general, not very practical.

Another common problem is reading the age of a star from its position in the HR diagram (or any other plot which does not show the age explicitly, e.g., the \textit{central temperature-central density} diagram).
It would be useful to have a quick access to that information and, at the same time, to know which of the many profiles describes the structure of the star at that age. 

Here we describe a graphical interface, MESAFace, which we created to address problems like the ones described above and, in general, to provide fast and simple access to \textit{all the information} contained in the output of a MESA run, including all the profiles.

MESAFace is a very intuitive, efficient, and easy-to-use dynamical interface, which allows the selection of the information to be shown through buttons and menus and the age of the star to be selected through a slidebar.

The code is written in Mathematica and has been tested with Mathematica 7 and 8.
However, a knowledge of Mathematica is not strictly necessary to use the interface.

MESAFace needs to access the MESA output data which, in a standard MESA run, are saved in the folder /work/LOGS.
The data can also be downloaded directly from the web.
Possibly soon, the results from some standard runs will be available from a dedicated  web space. 

After the data has been loaded, MESAFace presents a graphical interface structured in three vertically organized panels: 
\begin{itemize}
	\item A history plot panel;
	\item A profile plot panel;
	\item An information panel.
\end{itemize}
All the \textit{controls} (buttons, menus, check-boxes, etc.) are contained on the left side of the interface.
The user can decide what to plot and can change the age of the star by using these controls while the information shown on the right side is dynamically updated.
Several aspects of the panels and graphs, for example, the color and style of the lines, can be customized by the user.
A separate user manual~\cite{userManual} explains in detail how to do that.

In this paper, we describe in detail the structure of the MESAFace code and how it accesses, manipulates, and presents the data. 

The structure of the paper is the following:
We begin with a brief summary of the MESA output, in section \ref{sec:BriefDescriptionOfTheMESAOutput}.
Later, in section \ref{sec:TheStructureOfMf}, we give a basic description of the general structure of the MESAFace code.
In section \ref{sec:DownloadAndOrganizationOfTheFiles}, we describe with more details how MESAFace loads and organizes the files. 
In section \ref{sec:TheInterface}, we discuss how it creates the dynamical interface and visualizes the data.  
Finally, in section \ref{sec:Conclusions} we conclude with some overall comments. 

A note on the style. We will use the \texttt{Typewriter} style for the content of the Mathematica notebook, "quotes" for the name of the MESA files, and \textbf{bold face} for the names of the controls in the interface.

\section{Brief description of the MESA output}
\label{sec:BriefDescriptionOfTheMESAOutput}

MESA is a one-dimensional stellar evolution code.
The output data is parameterized by the value of the star age and the radial distance from the star center. 
Most of the output data is organized in a large number of \textit{profiles} ("log$n$.data" files, with $n$ an integer). 
The main section of each profile is a table with data referring to a particular age and each row referring to a particular \textit{zone} in the star. 
Different zones correspond to different distances from the center.

Besides the profiles, MESA also produces a "star.log" file which includes global information about the star at different times, and a "profiles.index" which provides a useful connection between the profiles and "star.log".
All these files are created during a MESA run and, by default, are contained in the LOG folder of the star/work directory.
Below is a brief description of the MESA output files. For more information, cfr. the MESA literature\cite{Paxton:2010ji,mesa}.

The MESA output folder contains 1 file "star.log", a large number of \textit{profiles}, 1 file, "profiles.index".
\begin{itemize}
	\item The file "star.log" includes global information about the star at different times.
	The second and third row of this file include the names (row 2) and values (row 3) of some physical quantities characterizing the star, for example its initial mass and metallicity. 
This information is accessible from the interface by selecting the \textbf{GENERAL} radio button in the GENERAL INFO panel of the interface (see section \ref{sec:Visualization of the data}).
	
	The sixth row contains the names of some \textit{age-dependent} physical quantities, such as effective temperature or central density, and all the rows below that show the values of these quantities at different ages. 
	Each row is indexed by a different \textit{model\_number} and represents a different age of the star.
	\item 	The \textit{profiles} ("log$n$.data" files) contain detailed information about the structure of the star, each file referring to a specific age. 
	Here, $n$ is the \textit{log file number}, also referred to as the \textit{log} or \textit{profile number}.

	The second and third row of these files include the names (row 2) and values (row 3) of some physical quantities characterizing the particular profile, for example the age of the star. 
	This information is accessible from the interface by selecting the \textbf{DETAILED} radio button in the GENERAL INFO panel of the interface (see section \ref{sec:Visualization of the data}).

	The sixth row contains the names of some \textit{position-dependent} physical quantities, such as temperature or density, and all the rows below that show the values of these quantities in the different zones. 
	\item The file "profiles.index" is an index of the "log$n$.data" files.
It has three columns. 
The most interesting (for our purpose) are the first and the third. 
The first contains the \textit{model\_number}, which indicates a specific row of the file "star.log" (each row has a different \textit{model\_number}).
The third column has the \textit{log file number}. 
In this way "profiles.index" provides a link among the various "log$n$.data" and "star.log".
\end{itemize}
The number of different profiles and the amount and kind of information in each file depend on the particular run. 

\section{General structure of the MESAFace code}
\label{sec:TheStructureOfMf}

The MESAFace code contains two scripts, labeled   as \texttt{(* SCRIPT 1~*)} and \texttt{(* SCRIPT 2~*)}, which should be run consecutively. 

The first script extracts all the data from a MESA run and organizes them into tables easily accessible by the interface (see section \ref{sec:DownloadAndOrganizationOfTheFiles}).
Although extracting all the MESA output at once may be slow, after that, the interface can access all the data very quickly, allowing a dynamical manipulation and visualization of the information.

The interface is created by the second script, labeled \texttt{(* SCRIPT 2~*)} in the code and described in section \ref{sec:StructureOfTheSecondScript} and \ref{sec:TheInterface}.

The choice to have two scripts is a practical one:
The second script, which normally runs in no more than a second or so, allows several customizations of the interface, for example it allows the user to set the style of the graphs.
Having this script separated from the first one allows the user to change some features in the interface or in the plot style without having to download the data each time. 

\subsection{Structure of the first script}
\label{sec:StructureOfTheFirstScript}

The first script can be divided in two sections: 
\begin{enumerate}
	\item The first section, which ends before the definition of the function \texttt{age}, sets the path to the folder with the MESA output data and downloads the data.
	\item The second section organizes the data (see section \ref{sec:DownloadAndOrganizationOfTheFiles}).	
\end{enumerate}
The variable \texttt{slash} allows Mathematica to read the location of the output files with different operating systems.
If the output data are on the web, the variable \texttt{slash} has to be set equal to \texttt{"/"}.

\subsection{Structure of the second script}
\label{sec:StructureOfTheSecondScript}

The second script can be divided into three main sections: 
\begin{enumerate}
	\item The section contained within the comment lines  
\texttt{(* General Settings~*)}
and 
\texttt{(* DON'T CHANGE~*)}
at the beginning of the second script allows the user to customize the interface. 
In this section, it is possible to define the style of the graphs, the labels, and the buttons to show in the interface. 
\item The section below the comment line \texttt{(* Plot and Info Section*)} produces the information to be presented in the three vertical panels on the right hand side of the interface (see section~\ref{sec:Visualization of the data}).
\item The section after the comment line \texttt{(* Image Size~*)} creates and manages the controls to be shown on the left hand side of the interface (see section~\ref{sec:DescriptionOfTheControlsVariables}). 
\end{enumerate}

\subsection{Local and global variables}
\label{sec:LocalAndGlobalVariables}

Most of the variables used in the scripts are local (they are visible only within the script).
A variable is made local by including it in the curly brackets at the beginning of the \texttt{Module} command.

However, some variables defined in the first script need to be accessed by the second script and, therefore, they cannot be defined locally. 
These global variables are:
\begin{itemize}
	\item \texttt{path}: it defines the path to the folder with the MESA output data.
	\item \texttt{star}: it is a table and contains the content of the file "star.log".
	\item \texttt{log[logNumber]}: it is an array of tables with the content of the profiles.
\texttt{logNumber} is an index equal to the \textit{log file number} (see section \ref{sec:BriefDescriptionOfTheMESAOutput}).
The table \texttt{log[n]} corresponds to the file "log$n$.data".
	\item \texttt{profiles}: it is a table and contains the content of the file "profiles.index".
	\item \texttt{logAge}: it is a two column table with, respectively, the log file number and the star age of each profile.
	\item \texttt{maxModelLength}: it is an integer equal to the number of lines of the profile which has the largest number of lines.
\end{itemize}
If it is necessary to debug or to use a result outside a MESAFace session, a local variable can be made global by taking it out of the list in the curly brackets at the beginning of the \texttt{Module} command. 

In general, however, it is not recommended to use nonlocal variables outside MESAFace during a session.

\section{Loading and organizing of the files}
\label{sec:DownloadAndOrganizationOfTheFiles}

The goal of MESAFace is to provide a way to access the data \textit{dynamically}. 
In order to do that, MESAFace downloads all the data at once, either from the hard-drive or from the web, and saves it in particular variables.
This may be a rather slow process, especially if the data are to be extracted directly from a web location.
After the data is downloaded, however, the access to the information is very fast. 

The progress in downloading is indicated: a numerical value represents the actual number of downloaded profiles while a progress-bar shows the relative progress.

\subsection{Extraction of the MESA output}
\label{sec:ExtractionOfTheMESAOutput}

The job of the first script of MESAFace is just to download and organize the MESA output data.
Mathematica looks in the folder defined by the variable \texttt{path} for all the files with names "star.log", "profiles.index" and "log$n$.data" (it is essential that the output files have those names, as it is for a default run).
The variable \texttt{path} can point to a folder on the hard drive or on the web.

The files "star.log" and "profiles.index" are saved (as tables) respectively 
in the variable \texttt{star} and \texttt{profiles}.
Analogously, the "log$n$.data" files are saved in the array of tables \texttt{log[indx]}, where \texttt{indx} is the \textit{log number} $n$.
For example, the file "log$33$.data" is saved as \texttt{log[33]}.

\subsection{Extraction of  the age of the star and selection of the profile to show}
\label{sec:SelectionOfProfileToShow}

One of the main features of MESAFace is to allow an easy selection of the profile corresponding to a certain age.
To do that, the first script organizes the profiles in order of age in the 
table \texttt{logAge}, with the \textit{log number} in column 1 and the star age in column 2.

The function \texttt{age}: 
\[
	\begin{array}{ll}  
& \texttt{age[i\_]:= log[i][[1 + Position[log[i], "star\_age"][[1, 1]],} \\ 
& \texttt{Position[log[i], "star\_age"][[1, 2]]]];}
	\end{array}       
\]
has the task of extracting the age of each profile.

A \textit{slide-bar} (more precisely, a \textit{manipulator}) type control:
\[
	\begin{array}{ll}  
& \texttt{\{\{starAge, logAge[[1, 2]], Style["Star Age", 12, \{Bold, Blue\}]\}, }\\
&\texttt{ logAge[[All, 2]], ControlType -> Manipulator\},}
	\end{array}       
\]
allows the user to set the value of the real variable \texttt{starAge} to any of the lines in the second column of the table \texttt{logAge} (which, as explained, contains the ages of all the profiles).

Each time \texttt{starAge} is selected, Mathematica uses the table \texttt{logAge} to find the corresponding \textit{log file number}, and saves it in the variable \texttt{model}:
\[
	\begin{array}{ll}  
& \texttt{model = logAge[[Position[logAge[[All, 2]], starAge][[1, 1]], 1]];}
	\end{array}       
\]
In addition, if the check-box \textbf{Show Age in History Plot} is checked, MESAFace finds which line of the file "star.log" corresponds to the age selected:
\[
	\begin{array}{ll}  
& \texttt{modelNumber =   profiles[[Position[Drop[profiles, 1][[All, 3]], }\\&\texttt{model][[1, 1]] + 1,    1]];}\\
&\\
& \texttt{indexAge = Position[Drop[star, 6][[All, 1]], modelNumber]}\\&\texttt{[[1, 1]];}	
\end{array}       
\]
Here, \texttt{modelNumber} is equal to the \textit{model number} discussed above (see section \ref{sec:BriefDescriptionOfTheMESAOutput}).
Its value is reported in the first column of "profile.index" and in the first column of "star.log".

The variable \texttt{indexAge}, instead, is equal to the the row number of "star.log" (after its first 6 lines are dropped) corresponding to the \textit{model number} (and therefore the age of the star).
This value is used to identify the position of the age of the current profile in the history plots, providing a link between the \textbf{History Plots} and \textbf{Profile Plots} panels.

\section{The interface}
\label{sec:TheInterface}

Running the second script produces a graphical interface in which the controls, located on the right hand side, allow the user to choose what kind of information to show in the visualization panels to the right. 

\subsection{Description of the controls}
\label{sec:DescriptionOfTheControlsVariables}

The controls are coded in the section of the second script which follows the comment line \texttt{(* Beginning of the proper dynamical part *)}.
They are described by variables and are normally nested in double curly brackets. 
For example, the line:
\[
	\begin{array}{ll}  
  & 	\texttt{\{\{variable name, Default value, "label"\}, list of possible values,	} \\
  & \texttt{ControlType -> PopupMenu\},}
	\end{array}     
\]
describes a variable whose value can be set by a popup-menu~\cite{manipulate}.


The first controls in each of the three sections of the interface are \textit{radio button}-type controls 
which allow the user to choose the kind of plot (top two panels) or information (bottom panel) to show.
They are coded in the section of the script below the comment line 
\texttt{(*~Beginning of the proper dynamical part~*)}, at the beginning of
the \texttt{(*~History  Plot  Section~*)}, the \texttt{(*~Profile  Plot  Section~*)}, and the \texttt{(*~Info  Section~*)}:
\[
	\begin{array}{ll}  
  &\texttt{\{\{plotType, "Customized", Style["Type of Plot", 12, Bold]\},}\\
  &\texttt{\{"Customized", "HR", "CentralAbundances", "Burnings"\},}  \\
  &\texttt{ControlType -> RadioButtonBar\},}\\
	&\\    
  &\texttt{\{\{profileType, "Custom", Style["Type of Plot", 12, Bold]\},}\\
  &\texttt{\{"Custom", "Abundances", "Reactions"\},  }\\
  &\texttt{ControlType -> RadioButtonBar\},}\\
	&\\    
  &\texttt{\{\{Info, "PATH", ""\},\{"PATH", "GENERAL", "DETAILED"\},}\\
  &  \texttt{ControlType -> RadioButtonBar\},}
	\end{array}  
\]
These controls allow the user to set the \textit{string} value of \texttt{plotType}, \texttt{profileType}, and \texttt{Info}.

When these variables are set at their default values, the quantities to be plotted are represented by the string-valued variables
\texttt{xH} and \texttt{yH}, for the $x-$ and $y-$coordinates of the history plots
and \texttt{xP} and \texttt{yP}, for the $x-$ and $y-$coordinates of the profile plots.

The possible values of \texttt{xH} and \texttt{yH} are the elements in the $6^{\rm th}$ row of the file "star.log".
The possible values of \texttt{xP} and \texttt{yP} are the elements in the $6^{\rm th}$ row of any of the files "log$n$.data".
All these options are accessible to the user through popup menus and buttons. 
The lists of buttons to show in the interface are coded in the \texttt{(*Buttons to show*)} section of the code:
\[
	\begin{array}{ll}  
&\texttt{(*Buttons to show*)} \\
& \texttt{HistoryButtonsX = \{"star\_age", "log\_Teff", "log\_L", "log\_R", }\\
&   \texttt{"log\_center\_T", "log\_center\_Rho"\};}\\
& \texttt{HistoryButtonsY = \{"log\_Teff", "log\_L", "log\_R", "log\_center\_T",} \\
&   \texttt{"log\_center\_Rho"\};}\\
&\texttt{HistoryButtonsY2 = \{"log\_center\_P", "center\_h1", "center\_he4",} \\
&   \texttt{"center\_c12"\}};\\
&\\
&\texttt{ProfileButtonsX = \{"mass", "radius", "logR"\};}\\
&\texttt{ProfileButtonsY = \{"radius", "mass", "logR", "logT", "logRho",}\\
&    \texttt{ "logP", "h1", "he3", "he4"\};}\\
&\texttt{ProfileButtonsY2 = \{"c12", "n14", "o16", "log\_opacity", } \\
&   \texttt{"luminosity", "non\_nuc\_neu"}\};
 \end{array} 
\]
The first element of each list gives the default value of the corresponding variable.

The plot functions \texttt{Customized} and \texttt{Custom}, plot the values in the two columns defined by  \texttt{xH} and \texttt{yH} of the file "star.log"
(after dropping the first 6 rows) and the values in the two columns defined by  \texttt{xP} and \texttt{yP} of the current file "log$n$.data" (again, after dropping the first 6 rows) respectively.

The other plot functions show specific columns of the files "star.log" or "log$n$.data" in order to produce some standard plots (see section \ref{sec:Visualization of the data}).

Regardless of the value of the \texttt{plotType} and \texttt{profileType} variables, the following parameters define some characteristics of the graphs:
\begin{itemize}
	\item \texttt{Cx}, \texttt{Cy}, \texttt{Dx}, \texttt{Dy} represent the scale factors for, respectively, the $x-$ and $y-$axis of the history plots and the $x-$ and $y-$axis of the profile plots.
	Selecting $n$ in one of the \textbf{axis scale} menu rescales the axis by a factor of $10^{-n}$.
For example, if the $x$-axis represents the age in years, selecting \texttt{Cx=6} changes the units to Myr.
	\item The string-valued variables \texttt{funcXH}, \texttt{funcYH}, \texttt{funcXP}, and \texttt{funcYP} define the operation to be performed on the data represented by, respectively, \texttt{xH}, \texttt{yH}, \texttt{xP} and \texttt{yP}. 
	Their possible value is selected by popup menu type controls. 
	The expression corresponding to the string is specified in the list \texttt{DataOperations}, in the section \texttt{(*~Operations on the data*)}.
		\item The real-valued variables \texttt{offsetXH}, \texttt{offsetYH}, \texttt{offsetXP}, and \texttt{offsetYP} define new zeros for the axes. 
	\item \texttt{startH} and  \texttt{endH} define the plot range for the history plots, that is, the rows of the file "star.log" to be shown.
Changing their value allows to zoom in a particular region of the history plot.	
	\item \texttt{startP} and  \texttt{endP} define the plot range for the profile plots, that is, the rows of the current profile to be shown.
Changing their value allows to zoom in a particular region of the profile plot.	
\end{itemize}
All the possible values for these variables can be set from the control section, on the left hand side of the interface.

The \texttt{(*~ Profile  Plot  Section*)} contains two more variables: \texttt{starAge} and \texttt{goToAge}.
\begin{itemize}
	\item \texttt{starAge} can be equal to the age of any of the profiles from the current simulation. 
Its value can be changed dynamically through a slide-bar and its current value is indicated on top of the profile plot.
Changing \texttt{starAge} corresponds to selecting a new "log$n$.data" file. 
The log number corresponding to the current age is also shown on top of the profile plot.
When the star age is changed, all the plots and the other information shown are dynamically updated.
	\item \texttt{goToAge} is a logical variable, by default equal to \textit{false}.
If its value is set to \textit{true} (by checking the box "Show Age in History Plot" in the profile section of the interface),
the age selected will be shown in the history plot through a line or a dot (or both, depending on the kind of plot).
The style of the line and of the dot is set by the parameters \texttt{AgeLineColor}, \texttt{AgeDotSize} and \texttt{AgeDotColor} in the 
\texttt{(*Graphs colors~*)} section of the script:
\end{itemize}
\[
	\begin{array}{ll}  
  & \texttt{AgeLineColor =} \{\texttt{Red, Dashed}\}\texttt{;}\\
  & \texttt{AgeDotSize = 0.01;}\\
	& \texttt{AgeDotColor = Red;}
	\end{array}       
\]

\subsection{Visualization of the data}
\label{sec:Visualization of the data}

When the second script is executed, Mathematica reads the value of the three \textit{string-valued} variables
\texttt{plotType}, \texttt{profileType}, and \texttt{Info}.

\begin{figure}[tp]
	\centering
		\includegraphics[width=0.4\textwidth]{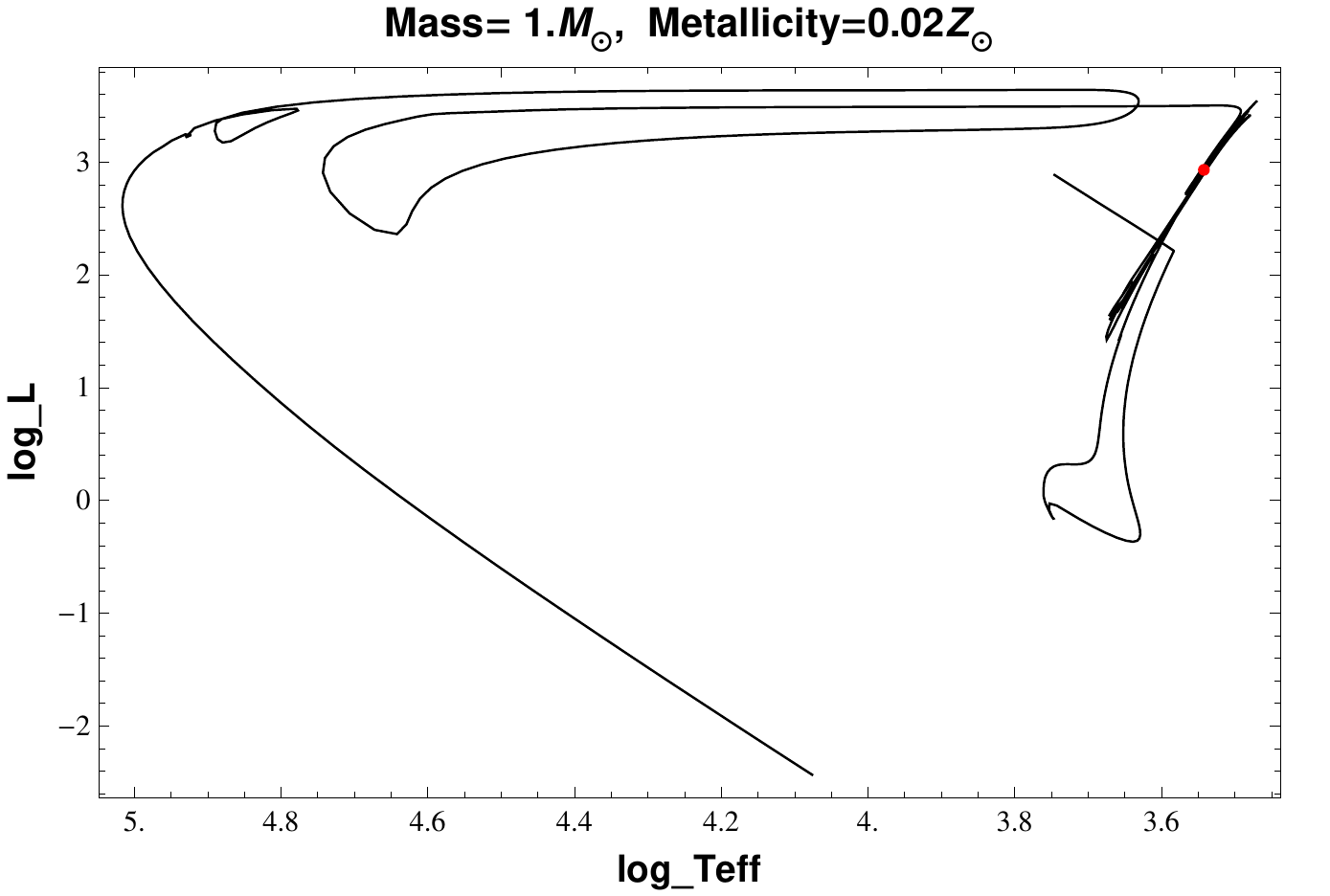}
		\includegraphics[width=0.4\textwidth]{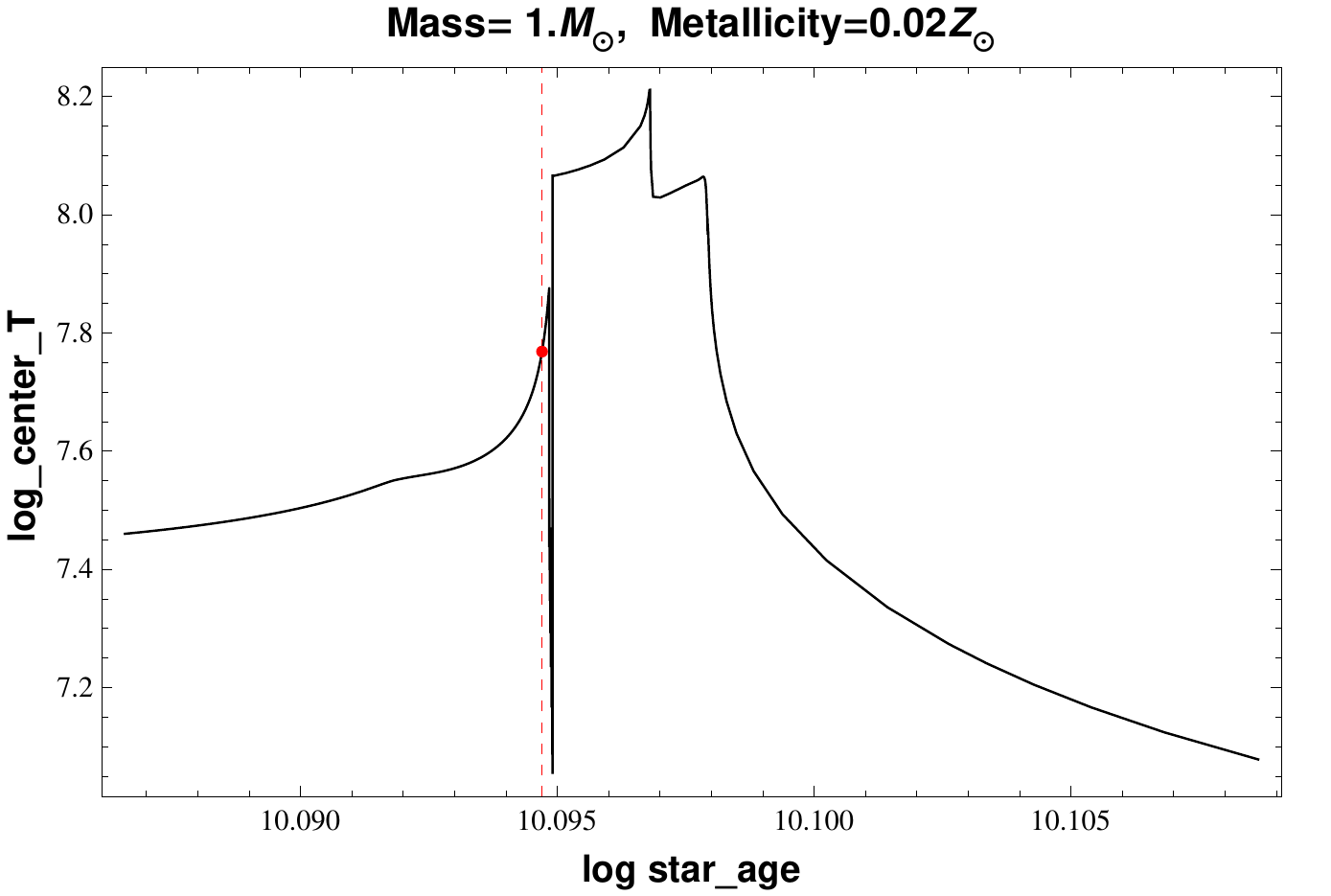}
		\includegraphics[width=0.4\textwidth]{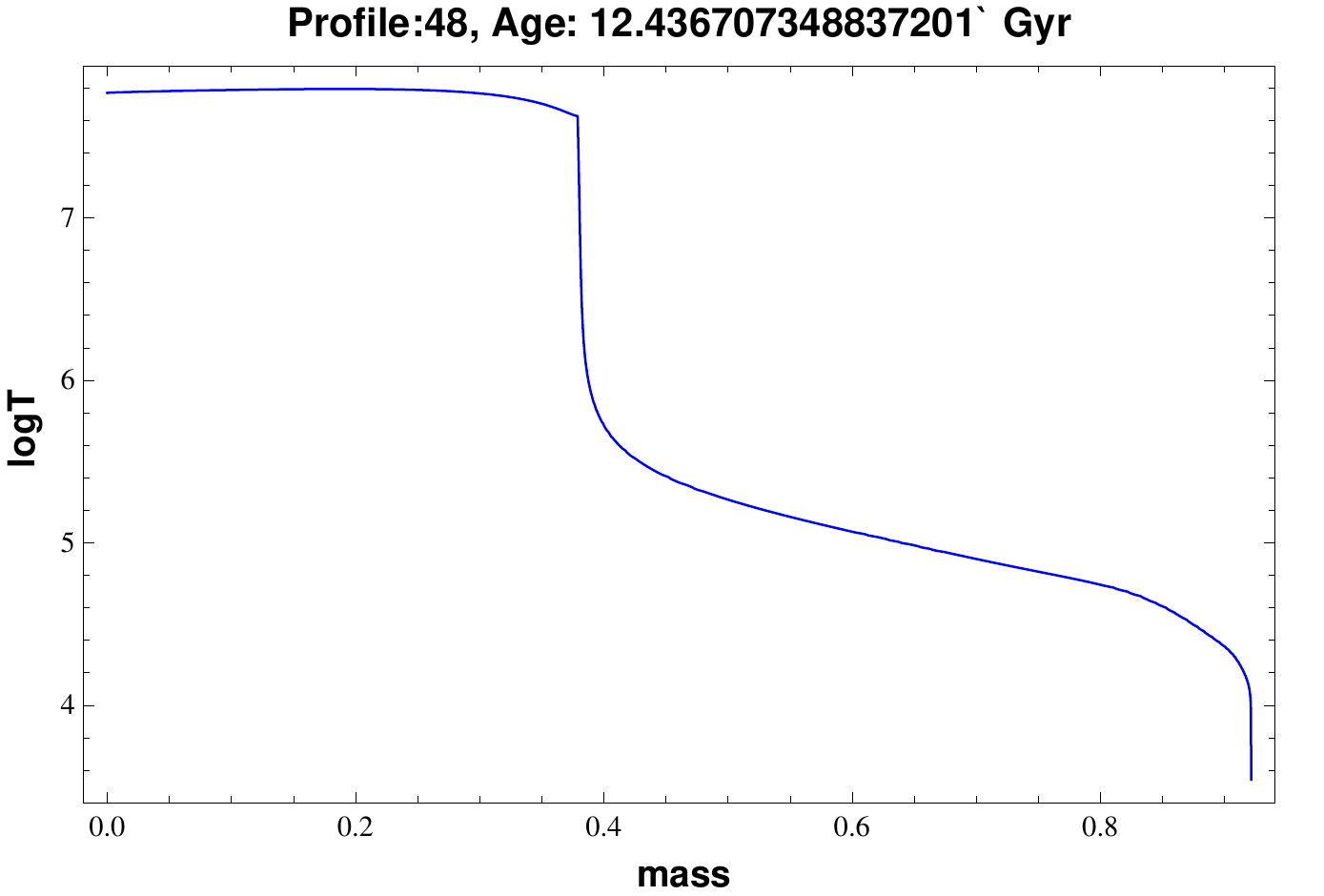}
		\includegraphics[width=0.4\textwidth]{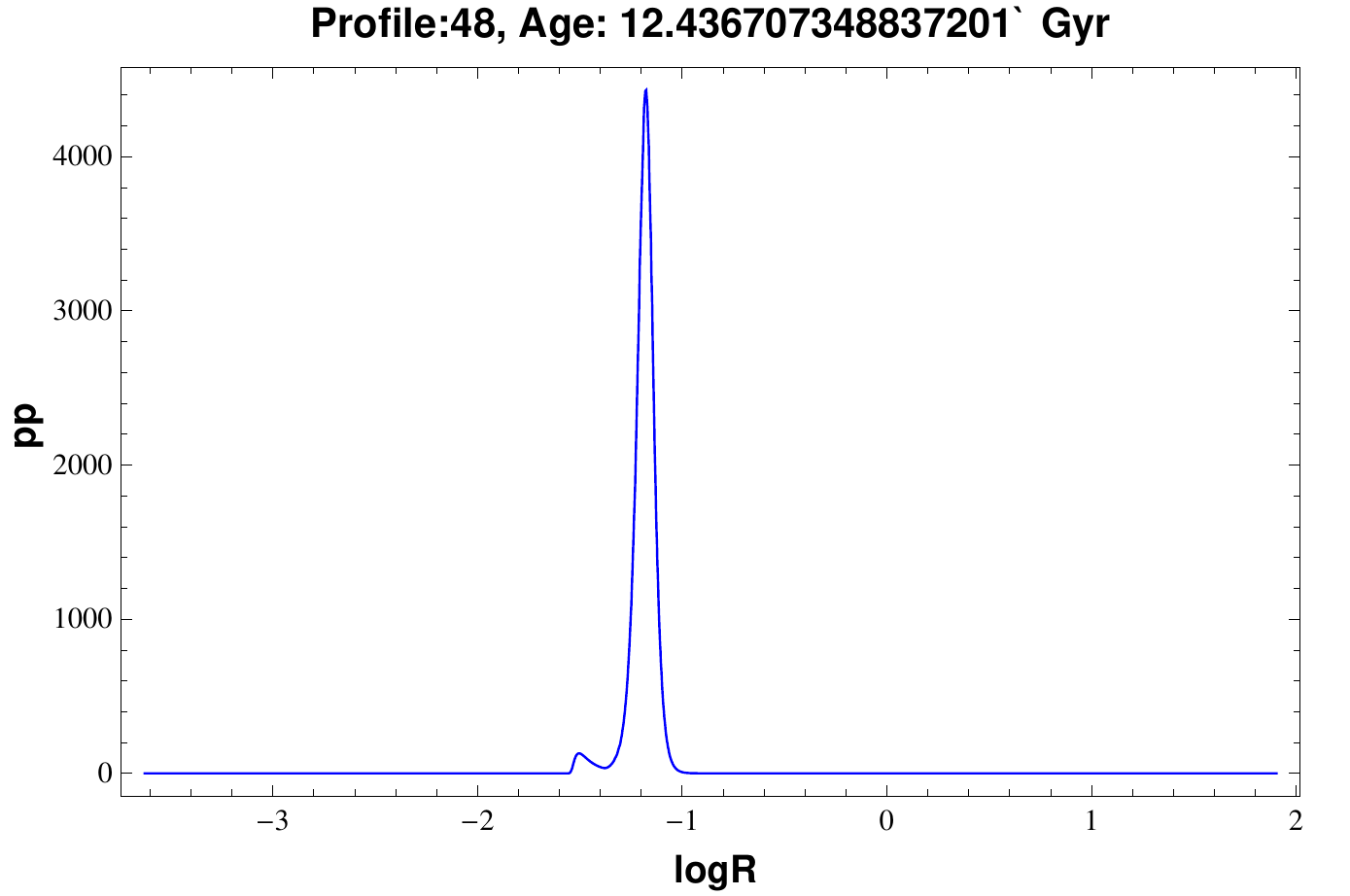}
	\caption{The plots refer to a 1 $M_\odot$ star, with solar metallicty~\cite{1MSun}.
	The top ones are examples of history plots. The bottom ones of profile plots. 
The 	red dot and the dashed line indicate the age corresponding to the profiles shown, in this case about 12.5 Gyr.
}
	\label{fig:graphs}
\end{figure}

Then, in the following section of the code: 
\[
	\begin{array}{ll}  
\texttt{Grid[}\{
  &\texttt{(*~History Plot ~*)}\\
  &\{\texttt{Dynamic[ToExpression[plotType]]}\},\\
  &\\
  &\texttt{(*~Profile Plot ~*)} \\
  &\{\texttt{Dynamic[ToExpression[profileType]]}\}, \\
  &\\
  &\texttt{(*~Information ~*)} \\
  &\{\texttt{Dynamic[ToExpression[Info]]}\}\}\texttt{,} \\
  &\texttt{Alignment} \to \texttt{Left, Frame} \to \texttt{All],}
	\end{array}
\]
these three strings are transformed into \textit{expressions} which are evaluated in the section of the code labeled \texttt{(*~Plot and Info Section*)}, described below.
In addition, the \texttt{Grid} function creates three panels which show the value of these three expressions.
Examples of graphs shown in the history and profile panels are given in Fig.\ref{fig:graphs}. 
%

The section \texttt{(*~Plot and Info Section*)}, 
is divided into three subsections, each responsible for one of the three graphical panels:
\begin{enumerate}
	\item \texttt{(*~History Plots Section~*)} for the top panel. 
	Defines the possible values for the variable \texttt{plotType}; 
	\item  \texttt{(*~Profile Plots Section~*)}  for the middle panel. 
	Defines the possible values for the variable \texttt{profileType}; 
	\item  \texttt{(*~Info Section~*)}  for the bottom panel. 
	Defines the possible values for the variable \texttt{Info}; 
\end{enumerate}

Below is the list of the possible values of \texttt{plotType}, with the results that they produce:
\begin{itemize}
	\item \textbf{Customized} is the default choice.
	In this case, the variables in the $x-$ and $y-$axes of the graphs (in the top panel) are selected from the buttons or from the \textit{popup} menus.
Each popup menu contains the names of all the variables that are possible to plot 
(i.e., the names of all the columns in the "star.log" file).
\end{itemize}

\begin{itemize}
	\item \textbf{HR} produces the HR diagram, with the logarithm of the effective temperature on the $x-$axis and the logarithm of the luminosity on the $y-$axis.
	The temperature increases toward the left of the $x-$axis, in line with the standard convention for the HR diagram.
\end{itemize}

\begin{itemize}
	\item \textbf{CentralAbundances} shows the central abundance of the elements defined in the list \texttt{CentralAbundancesToShow} in the \texttt{(* General Settings *)} section of the code.
\end{itemize}

\begin{itemize}
	\item \textbf{Burnings} shows the energy released in the nuclear reactions defined in the list \texttt{CentralAbundancesToShow} in the \texttt{(* General Settings *)} section of the code.

\end{itemize}

The possible values of \texttt{profileType} are 
\begin{itemize}
	\item \textbf{Custom} is the default choice and works in the same way as for the history plots.
\end{itemize}
\begin{itemize}
	\item \textbf{Abundances} shows the profile of the abundance of the elements defined in the list \texttt{ProfileAbundancesToShow} in the \texttt{(* General Settings *)} section of the code.
\end{itemize}
\begin{itemize}
	\item \textbf{Reactions} shows the energy released in the nuclear reactions defined in the list \texttt{ProfileBurningsToShow} in the \texttt{(* General Settings *)} section of the code.
\end{itemize}
All the lists for the abundances and the reactions can be customized according to the user's needs.

%
%

Finally, the possible values of \texttt{Info} are
\begin{itemize}
	\item  \textbf{PATH} is the default value.
In this case MESAFace shows the value of the variable \texttt{path} and so indicates the working folder or the web location from where Mathematica is extracting the data.
	\item  \textbf{GENERAL} shows general information about the star not related to the current profile. 
	The information is taken from the second and third rows of the file "star.log".
	\item  \textbf{DETAILED}  shows detailed information about the current profile.
	The information is taken from the second and third rows of the current profile.
\end{itemize}

It is possible to add other kinds of plots or information by adding new options for the variables \texttt{plotType}, \texttt{profileType}, and \texttt{Info} in the section below the comment line \texttt{(*~Beginning of the proper dynamical part~*)} and then define their \textit{expression-values} in the \texttt{(*~Plot and Info Section*)} of the code.

\section{Conclusions}
\label{sec:Conclusions}

MESAFace is a user-friendly and efficient interface to analyze the MESA output data. 
It is written in Mathematica, a very powerful and very well-documented mathematical software, and it is easily customizable.

The main purpose of this interface is to provide a tool to quickly analyze all the output data from a MESA run through graphs and tables.
Our final objective was to have a very intuitive and easy-to-use instrument, accessible to researchers as well as to students.
MESAFace allows the user to choose what to plot through various menus and to scroll the various profiles, organized by age, with a slidebar.
The user can also easily zoom in different regions of the plot, change units, set a different zero for each axis, or plot the log (among other functions) of the data.

The production of high-quality graphs was not our priority, as other tools are available~\cite{tioga}.
However, improving the quality and create new options to manipulate the aspect of a graph, 
is a prominent direction for future improvement of MESAFace.

This paper is not meant to be a proper user manual. 
Basic instructions on how to use MESAFace are accessible on-line~\cite{userManual}. 
Here, instead, we have described in detail the structure of the code and the meaning of the variables and controls.
This allows the user not just to use, but if necessary, to modify and possibly improve this code.

\section*{Acknowledgments}

We express our gratitude to Dr. Stan Owocki, who encouraged us to develop this interface.

%




%

\begin{thebibliography}{99}
\bibitem{manipulate} 
Many more details on the controls can be found in the Mathematica help for \texttt{Manipulate}.
\bibitem{1MSun} 
	The inlist file to generate the data is available in the MESA test suite folder of MESA. 
The resulting data folder can also be downloaded directly from~\cite{userManual}.
\end{thebibliography}







\end{document}